\documentclass[aps,pre,twocolumn,showpacs,superscriptaddress,10pt]{revtex4}
\usepackage{color}
\usepackage{graphicx}
\usepackage{amsmath}
\usepackage{amsfonts}
\usepackage{amssymb}
\usepackage{amsthm}
\usepackage[colorlinks,citecolor=blue,linkcolor=blue]{hyperref}

\begin{document}

\title{Classical heat transport in anharmonic molecular junctions: exact solutions}

\author{Sha Liu}
\email{liusha@nus.edu.sg}
\affiliation{Department of Physics and Centre for Computational Science and
Engineering, National University of Singapore, Singapore 117546}
\affiliation{NUS Graduate School for Integrative Sciences and Engineering,
Singapore 117456}
\author{Bijay Kumar Agarwalla}
\affiliation{Department of Physics and Centre for Computational Science and
Engineering, National University of Singapore, Singapore 117546}
\author{Jian-Sheng Wang}
\affiliation{Department of Physics and Centre for Computational Science and
Engineering, National University of Singapore, Singapore 117546}
\author{Baowen Li}
\email{phylibw@nus.edu.sg}
\affiliation{Department of Physics and Centre for Computational Science and
Engineering, National University of Singapore, Singapore 117546}
\affiliation{NUS Graduate School for Integrative Sciences and Engineering,
Singapore 117456}
\affiliation{Center for Phononics and Thermal Energy Science, School of
Physical Science and Engineering, Tongji University, 200092 Shanghai, China}


\date{14 Jan 2012}
\newcommand{\ddd}[1]{{#1}^{\circ}}
\newcommand{\pt}[2]{\partial_{#1}{#2}}
\newcommand{\ppt}[3]{\partial^{#1}_{#2}{#3}}
\newcommand{\lpt}[2]{\frac{\partial{#1}}{\partial{#2}}}
\newcommand{\lptl}[1]{\lpt{#1}{\lambda}}
\newcommand{\ptl}[1]{\pt{\lambda}{#1}}
\newcommand{\pptl}[1]{\partial^2_{\lambda}{#1}}
\newcommand{\lppt}[3]{\frac{\partial^{#1}{#2}}{\partial{#3}^{#1}}}
\newcommand{\eeqref}[1]{Eq.~(\ref{#1})}
\newcommand{\bkb}[1]{\left(#1\right)}
\newcommand{\bks}[1]{\left[#1\right]}
\newcommand{\eg}[1]{{\it e.g.}}
\newcommand{\ie}[1]{{\it i.e.}}
\newcommand{\fracc}[2]{(#1)/(#2)}
\newcommand{\tdl}{\tilde{L}}
\newcommand{\tdm}{\tilde{\mu}}
\newcommand{\tdv}{\tilde{V}}
\newcommand{\tdps}{\tilde{\Psi}}
\newcommand{\ttdps}{{\Psi}}
\newcommand{\ttdvp}{\phi}
\newcommand{\ttdp}{\varphi}
\newcommand{\tdp}{\tilde{\ttdp}}
\newcommand{\tdvp}{\tilde{\ttdvp}}
\newcommand{\tdt}{\tilde{T}}
\newcommand{\tdz}{\tilde{z}}
\newcommand{\tdZ}{\tilde{Z}}
\newcommand{\tdc}{\tilde{c}}
\newcommand{\Pst}{P^{\mathrm{st}}_T}
\newcommand{\tdg}{\tilde{G}}
\newcommand{\mmean}[1]{\left\langle\!\left\langle#1\right\rangle\!\right\rangle}
\newcommand{\meant}[1]{\left\langle#1\right\rangle_T}
\newcommand{\meantt}[2]{\left\langle#1\right\rangle_{T=#2}}
\newcommand{\mean}[1]{\bar{#1}}
\newcommand{\eval}[1]{\left.#1\right|_{\lambda=0}}
\newcommand{\inner}[2]{\left(#1,#2\right)}
\newcommand{\iinner}[2]{(#1,#2)}
\newcommand{\mathsp}[1]{\quad\mathrm{#1}\quad}
\newcommand{\lpta}[1]{\lpt{#1}{T_1}}
\newcommand{\lptb}[1]{\lpt{#1}{T_2}}
\newcommand{\pta}[1]{\pt{T_1}{#1}}
\newcommand{\ptb}[1]{\pt{T_2}{#1}}
\newcommand{\dif}[1]{\eval{\lptl{#1}}}
\newcommand{\Py}{P_y}
\newcommand{\Pq}{P_Q}
\newcommand{\mytextcolor}[2]{#2}

\begin{abstract}

We study full counting statistics for classical heat transport through {\it anharmonic/nonlinear}
molecular junctions formed by interacting oscillators.
Analytical result of the steady state heat flux for an overdamped anharmonic
junction with arbitrary temperature bias is obtained. It is found that the thermal
conductance can be expressed in terms of temperature dependent effective force
constant. The role of anharmonicity is identified.
We also give the general formula for the second cumulant of heat in steady state, as well
as the average geometric heat flux when
two system parameters are modulated adiabatically. We present an
anharmonic example for which all cumulants for heat can be obtained exactly. For a bounded
single oscillator model with mass we found that the cumulants are independent of
the nonlinear potential.
\end{abstract}

\maketitle

\newcommand{\mysec}[1]{\section{#1}}

\section{Introduction}
 Anharmonicity or nonlinearity plays an important role in many physical
processes. For example, it is crucial for umklapp scattering which
gives rise to a finite thermal conductivity for bulk systems as pointed
out by Peierls \cite{NULL.55.Peierls}. It is also an essential ingredient
required for developing functional phononic devices
\cite{RMP.12.Li}, such as thermal rectifier \cite{PRL.02.Terraneo, S.06.Chang} and heat
pump \cite{PRE.06.Segal,PRL.08.Segal,PRL.10.Ren}.


%
%

Comprehensive understanding of the general behaviors of nonlinearity in
transport is a fundamental problem in condensed matter physics and nonequilibrium
statistical physics.
Although elegant approaches exist to deal with ballistic transport in
harmonic systems \cite{JSP.06.Dhar,EPJB.08.Wang},
solving an anharmonic model analytically is extremely challenging
due to the intrinsic non-integrability of \mytextcolor{blue}{equations of
motion}. As a result,
for classical heat transport, the major approaches are limited to molecular
dynamics simulations \cite{PR.03.Lepri,AP.08.Dhar}, perturbation theories such as Boltzmann
equation \cite{NULL.55.Peierls,PRE.03.Pereverzev}, mode coupling theory
\cite{EL.98.Lepri,JSM.07.Delfini}, and/or mean--field theory \cite{EL.06.Li,PRE.08.He}.
There does not exist analytic solutions of heat transport problem for general anharmonic
potentials.


Moreover, in heat transport problems, not only the heat \mytextcolor{blue}
{flux}, but also the higher order
fluctuations (noises) are of significant interest as they contain valuable
information about the transport properties and satisfy
universal symmetry known as
Fluctuation Theorems (FT) \cite{PRL.97.Jarzynski,PRE.99.Crooks,PRL.95.Gallavotti}.
Therefore, methods to obtain these higher order moments and to verify this
symmetry for arbitrary anharmonic potentials are also highly desirable.

%
In this article, we present a general approach to nonlinear heat transport by working
in an enlarged space, which includes the heat $Q$ as an extra variable in the Fokker-Planck
(FP) equation \cite{PRL.12.Ren}. The heat current can be expressed as the
expectation value of certain
operator. If the steady state of the problem is known, then the analytic
expression \mytextcolor{blue}{for the current} can be obtained. We show this is the case
for two different anharmonic oscillator systems,
{\it i.e.}, the overdamped interacting double-oscillator model
(Model I) and the bounded
single-oscillator model (Model II).

For Model I, the complete cumulants, as well as the geometric
contribution to the heat flux, can be obtained if the full spectrum of the original FP
equation is solvable. One \mytextcolor{blue}{such} example is the V-shaped potential
$V(y)=k|y|$. For Model II, the complete cumulants can always be analytically obtained no
matter what the nonlinear interacting potential is.
For both models, the
Gallavotti-Cohen (GC) symmetry can be verified for arbitrary anharmonic interactions.

In the rest of the article,  Sec. \ref{sec:md1} is devoted to Model I and Sec.
\ref{sec:md2} is devoted to Model II. After that, we briefly summarize the article in
Sec. \ref{sec:summary}.

\section{\label{sec:md1}Model I} 
\begin{figure}[t!]
\centering
 \includegraphics[width=0.35\textwidth]{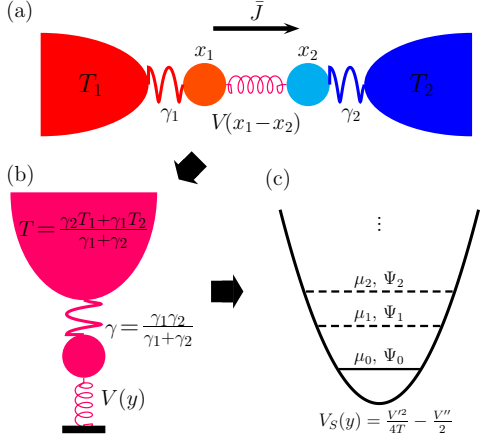}
 \caption{Schematic picture for the process of mathematical formalism for model I. (a)
The original problem: two coupled oscillators
connected to heat baths. (b) The reduced problem: single particle in contact with single
heat bath. (c) The ultimate problem: Schr\"odinger equation for a single particle.}
\label{fig:1}
\end{figure}

Model I consists of two coupled Brownian
oscillators which are connected to two Langevin heat baths (Fig. \ref{fig:1}(a)). The
dynamics is described by the Langevin
equations: $m_i
\ddot{x}_i= -\pt{x_i}V(x_1-x_2)-\gamma_i \dot{x}_i+ \xi_i$,
($i=1,2$), where $m_i$ and $x_i$ are the mass and displacement of oscillator $i$.
$\gamma_i$ is the coupling strength between the oscillator $i$ and the bath $i$, $T_i$ is
the temperature of bath $i$ and $\xi_i$ is a white noise with variance
$\left\langle\xi_i(t)\xi_j(t')\right\rangle= 2\gamma_iT_i\delta_{ij}\delta(t-t')$ ($k_B
$ is set to 1). $V$ depicts the interaction
potential, which is assumed to \mytextcolor{blue} {be bounded} and depend only on the
distance
between the two oscillators.

This model is a minimal model for heat transport in interacting oscillator systems. At low Reynolds number, we
can neglect the inertia and thus obtain the overdamped dynamics \cite{AJP.77.Purcell}:
\begin{align}
&\gamma_1\dot{x}_1+ V'(x_1-x_2)=\xi_1; \\
&\gamma_2\dot{x}_2-V'(x_1-x_2)=\xi_2,
\end{align}
in which the prime $'$ denotes the derivative with respect to its argument. The total heat
transferred from bath 1 to
bath 2 up to time $t$, denoted as $Q(t)$, satisfy
\begin{equation}
\dot{Q}=
V'(x_1-x_2) \dot{x}_1.
\end{equation}

By introducing \mytextcolor{blue} {the relative coordinate} $y= x_1-x_2$, we obtain two
stochastic
differential equations for the random variables $y$ and $Q$:
\begin{align}
	\dot{y}&= -\frac{V'(y)}{\gamma_1}-\frac{V'(y)}{\gamma_2}+
\frac{\xi_1}{\gamma_1}-\frac{\xi_2}{\gamma_2}; \\
\dot{Q}&=\frac{V'(y)}{
\gamma_1}\left(\xi_1-V'(y)\right).
\end{align}
Their joint probability distribution $\rho(y,Q,t)$ then evolves according to the
FP equation \cite{NULL.98.Risken,JSM.11.Kundu}
\begin{equation}
\label{eq:fk}
\begin{split}
	\lpt{\rho}{t}=&\left[
	\frac{T}{\gamma}\lppt{2}{}{y}+ \frac{1}{\gamma}\lpt{}{y}V'+
	\frac{T_1}{\gamma_1}\lppt{2}{}{Q}V'^2\right.\\
&\left.+\lpt{}{Q}\left(\frac{V'^2}{\gamma_1}-\frac{T_1V''}{\gamma_1}\right)
+\frac{2T_1}{\gamma_1}\frac{\partial^2}{\partial Q\partial y}V'
	\right]\rho,\\
	\end{split}
	\end{equation}
where $\frac{1}{\gamma}=\frac{1}{\gamma_1}+\frac{1}{\gamma_2}$ and
$\frac{T}{\gamma}=\frac{T_1}{\gamma_1}+\frac{T_2}{\gamma_2}$.

By taking the Fourier transformation on $Q$ in \eeqref{eq:fk}, we
obtain the equation for the characteristic function (CF)
$z(y,i\chi,t)=\int \rho(y,Q,t) e^{i\chi Q}dQ$, where $\int$ is short
for $\int_{-\infty}^{\infty}$ throughout the paper. Moreover, we introduce $\lambda\equiv
i\chi$ and use tilde ( $\tilde{}$ ) to express the $\lambda$-dependence (in such a
notation, $\lambda$ is
omitted in the parameter list). Then the equation reads
\begin{equation}
\label{eq:pdf}
\pt{t}{\tilde{z}(y,t)}= \tilde{L}\tilde{z}(y,t)
\end{equation}
with twisted FP operator
\begin{equation}
\label{eq:L}
\begin{split}
	\tilde{L}=& \frac{T}{\gamma}\lppt{2}{}{y}+ \frac{1}{\gamma}\lpt{}{y}V'+ \\
	&\lambda\bkb{\frac{T_1V''}{\gamma_1}-\frac{V'^2}{\gamma_1}-
\frac{2T_1}{\gamma_1}\lpt{}{y}V'}
+\lambda^2 \frac{T_1V'^2}{\gamma_1} \\
\equiv& L^{(0)}+ \lambda L^{(1)}+ \lambda^2 L^{(2)},
\end{split}
\end{equation}
For any quantity or operator $\tilde{\mathcal{A}}$, We shall also use $\mathcal{A}$ to be
short for
$\tilde{\mathcal{A}}$ evaluated at $\lambda=0$. For example,
$L=\tilde{L}(\lambda=0)=L^{(0)}$ and $z(y,t)=\int\rho(y,Q,t)dQ$ is the probability
distribution for $y$.
Moreover, for $\lambda=0$, the
equation $\pt{t}z(y,t)= Lz(y,t)$ actually \mytextcolor{blue}
{depicts} a single
Brownian oscillator connected with a single heat bath at \mytextcolor{blue} {an effective
temperature} $T$ and subjected to
a potential $V$ (see Fig.~\ref{fig:1}(b)). Therefore, the stationary solution will be
\mytextcolor{blue} {in the
form} of Boltzmann distribution
$
	z(y,t)\to\Pst(y)= \frac{1}{Z_{T}} e^{-V(y)/T}$ with $Z_{T}=\int
e^{-V(y)/T}dy
$ as the partition function.

We are only interested in the heat $Q$. Therefore, by integrating over $y$, we obtain the
CF of $Q$ as $\tilde{Z}(t)=\int
\tilde{z}(y,t)dy=\int \Pq(Q,t) e^{\lambda Q}dQ$, with
$\Pq(Q,t)=\int \rho(y,Q,t)dy$. The cumulant generating
function
(CGF) reads $\tdg\equiv\ln{\tdZ}$, which generates the $n$th-order
cumulant as $\mmean{Q^n}=
\eval{\partial_\lambda^n{\tdg}}$.

Obviously, if
we are able to solve \eeqref{eq:pdf} for general potential $V$, then the problem can be
completely solved, which is the case for harmonic potential \cite{PRL.12.Ren}.
For anharmonic cases,                    a general method is not available.

However, using separation of variables
$\tilde{z}(y,t)= \tdp(y)\tilde{F}(t)$,
\eeqref{eq:pdf} can still be casted into an eigenvalue problem
\begin{equation}
\label{eq:eigen}
	\tilde{F}(t)= e^{-\tilde\mu t},\quad\quad \tilde{L}\tdp(y)=-\tilde{\mu}\tdp(y).
\end{equation}
with natural boundary conditions for $y\to \pm \infty$. The structure of the solutions
is clearer if we introduce a transformation
$\tdp(y)=e^{-{V(y)}/{2\tilde T}}\tilde{\Psi}(y)$ with
$\tilde{T}=\fracc{\frac{T_1}{\gamma_1}+\frac{T_2}{\gamma_2}}{\frac{1}{\gamma_1}+
\frac{1}{\gamma_2}- \frac{2T_1}{\gamma_1}\lambda}$. By direct substitution, it turns out
that $\tilde\Psi(y)$ satisfy the following time independent Schr\"odinger equation
\begin{equation}
\label{eq:sdg}
\tilde L_S \tilde\Psi\equiv \frac{1}{\gamma}\bkb{- T
\lppt{2}{}{y}+ \tilde{V}_S}\tilde\Psi= \tilde\mu \tilde\Psi,
\end{equation}
which describes a single particle moving in the effective potential $\tilde V_S$ (see
Fig. \ref{fig:1}(c)), given by
\begin{equation}
\label{eq:sdg2}
	\tilde V_S=\frac{V'^2}{4T} \bks{1+
\frac{4T_1T_2\gamma}{\gamma_1+\gamma_2}
\lambda\!\left(\!
\frac{1}{T_1}\!-\!\frac{1}{T_2}-\lambda\!\right) }- \frac{1}{2}V''.
\end{equation}
Then the well-known results in quantum mechanics for single
particle can be directly evoked to discuss our situation. Two of them are that
all the eigenvalues are real (regard $\lambda$ as a real quantity) which can be ordered as
$\quad \tdm_0\le \tdm_1 \le \cdots$,
and upon normalization, all the eigenfunctions form a complete orthonormal basis set,
which satisfy
$
\label{eq:bset1}
	(\tdps_n,\tdps_m)=\delta_{nm}$ and $\quad\sum_{n=0}^{\infty} \tdps_n(y)\tdps_n(y')=
\delta(y-y').
$
We use $(f,g)$ to denote the inner product $\int f(y)g(y)dy$.

The original operator $\tdl$ is not Hermitian, \ie{}, $\tdl^\dag\neq \tdl$. However, it
can be shown that $\tdvp_n= e^{V/2\tdt} \tdps_n$ satisfy $\tdl^\dag \tdvp_n=
\tdm_n\tdvp_n$, namely, $\tdvp_n$ is an eigenfunction of $\tdl^\dag$ with eigenvalue
$\tdm_n$. When normalized properly, we will have
\begin{equation}
\label{eq:norm}
	(\tdvp_n,\tdp_m)=\delta_{nm},\quad\sum_{n=0}^{\infty} \tdvp_n(y)\tdp_n(y')=
\delta(y-y').
\end{equation}
Then we can write down the {formal} solution of \eeqref{eq:pdf} as
\begin{equation}
\label{eq:sol}
	\tdz(y,t)= \sum_{n=0}^{\infty} \tdc_n e^{-\tdm_n t}\tdp_n(y) ,
\end{equation}
where $\tdc_n$ is time independent and can be determined using the initial condition,
$\tdc_n= \inner{\tdvp_n}{\tdz(y,t=0)}$.


\subsection{Dynamic Heat Fluxes} Based on the formalism above, we come to discuss
the heat fluxes in nonequilibrium steady state. It should be pointed out that the discussions here are also
applicable to the dynamic heat fluxes \cite{PRL.10.Ren,PRL.12.Ren} when system parameters
are
modulated adiabatically.

If the long-time behavior is of main interest, then the
component with {smallest} eigenvalue ($\tdm_0$) will dominate, so that $\tdz(y,t)\to
\tdc_0
e^{-\tdm_0
t}\tdp_0(y) $ and $\tdZ(t)\to \tdc_0 e^{-\tdm_0 t}\int \tdp_0(y)dy$. Consequently, the
CGF $\ln{\tdZ(t)}\to -\tdm_0 t$ and the $n$th cumulant of the heat fluctuations are given
by
\begin{equation}
\label{eq:cum}
	\lim_{t\to \infty}\frac{\mmean{Q^n}}{t}=-\eval{\lppt{n}{\tdm_0}{\lambda}}.
\end{equation}
Of course, one would not be able to get all the cumulants unless  $\tdm_0$ can be obtained
as a function of $\lambda$. However, if one is only interested in the first
cumulant, \ie{}, the average heat flux $\mean{J}=
\lim_{t\to\infty}\mmean{Q}/t=-\eval{\pt{\lambda}{\tdm_0}}$, then here is an approach.

We differentiate \eeqref{eq:eigen} with respect to $\lambda$ and obtain
\begin{equation}
	(\ptl{\tdl})\tdp_0+ \tdl\ptl{\tdp_0}= -(\ptl{\tdm_0})\tdp_0- \tdm_0\ptl{\tdp_0}.
\end{equation}
Then we perform $\iinner{\tdvp_0}{\cdots}$ on both sides. Noticing that $\iinner{\tdvp_0}{
\tdl\ptl{\tdp_0}}= \iinner{\tdl^\dag\tdvp_0}{\ptl{\tdp_0}}=-
\tdm_0\iinner{\tdvp_0}{\ptl{\tdp_0}}$, we obtain
\begin{equation}
\label{eq:J}
	\mean{J}= -\eval{\ptl{\tdm_0}}=\eval{\inner{\tdvp_0}{(\ptl{\tdl})\tdp_0}}=
	\inner{\ttdvp_0}{L^{(1)}\ttdp_0}.
\end{equation}
This equation is the analogue of the Hellmann-Feynman theorem in quantum mechanics.
Note that $\phi_0$ and $\varphi_0$ in \eeqref{eq:J} are evaluated at
$\lambda=0$, therefore, are associated with the FP operator $L=L^{(0)}$. It can
be shown that the smallest eigenvalue for $L(L^\dag)$ is $\mu_0=0$ and
the corresponding eigenfunctions are
$
	\ttdp_0(y)\propto e^{-V(y)/T}\propto \Pst$ and $\ttdvp_0(y)\propto 1,
	$
respectively \cite{NULL.98.Risken}.

After normalization, \eeqref{eq:J} becomes
$
\mean{J}=
\int L^{(1)} \Pst(y)dy
$. In this sense, $L^{(1)}$ serves as the heat flux operator, which gives the average
heat flux when acted on the stationary state. Using $L^{(1)}=
\frac{1}{\gamma_1}(T_1V''-V'^2-
2T_1\lpt{}{y}V')$, we finally obtain
\begin{equation}
\label{eq:J2}
\mean{J}=  \frac{T_1}{\gamma_1} \meant{V''}-\frac{1}{\gamma_1}\meant{V'^2}=
\frac{\meant{V''}}{\gamma_1+\gamma_2}(T_1-T_2),
\end{equation}
where $\meant{\mathcal{A}}\equiv \int \mathcal{A} \Pst dy$ is the canonical
ensemble average of $\mathcal{A}$ at temperature $T$ and we have used
$\meant{V'^2}= T\meant{V''}$. Remember that $T=(\gamma_2 T_1+\gamma_1
T_2)/(\gamma_1+\gamma_2)$.

Equation (\ref{eq:J2}) is exact for a general potential $V$. It is also applicable to
the cases
with arbitrarily large temperature differences, {\it i.e.}, the cases in a
far-from-equilibrium state.
As far as we know, this is the first exact result for anharmonic heat transport with
general interaction, although it is
only for a small system with overdamped dynamics.

For harmonic potential $V=\frac{1}{2}ky^2$, $\meant{V''}= k$, which is just the
force constant and is independent of the temperature. For anharmonic potentials,
$\meant{V''}$ will generally be temperature dependent. In analogy with the
harmonic case,
we define $k_V(T)\equiv\meant{V''}$ as the effective force constant.

We notice that the temperature dependence of the effective force constant plays a key
role in many of the phenomena for heat control and management. For example,
a thermal rectifier is a device that directs the flow of heat \cite{PRL.02.Terraneo,
S.06.Chang}. To use this oscillator
system as a thermal rectifier, it is required that
\begin{equation}
	\left|\frac{\mean{J}(T_1=T_H,T_2=T_L)}{\mean{J}(T_1=T_L,T_2=T_H)}\right|=
	\frac{\meantt{V''}{\frac{\gamma_2 T_H+\gamma_1
T_L}{\gamma_1+\gamma_2}}}{\meantt{V''}{\frac{\gamma_2 T_L+\gamma_1
T_H}{\gamma_1+\gamma_2}}}\ne 1.
\end{equation}
Obviously, this condition is satisfied only when $k_V(T)$ is temperature dependent and the
system is asymmetric, namely, $\gamma_1\ne \gamma_2$.

As another example, in \cite{PRE.09.Li}, it was demonstrated that heat can be shuttled
across a symmetric system when the temperature of one heat bath is modulated
adiabatically but with zero average temperature bias, \eg{}, $T_1= T_2+ f(t)$
with $\overline{f(t)}=0$. If our system is modulated in such a way with
$\gamma_1=\gamma_2=2\gamma$ and $f(t)$ being a periodic function of $t$, one can obtain an
average heat flux
$
	\mean{J}=\frac{1}{4\gamma T_p}\int_0^{T_p}
k_V\!\!\bkb{\!\frac{2T_2+f(t)}{2}\!}\ f(t) dt
$
in one period. Again, $\mean{J}$ is nonzero only if $k_V(T)$ depends on $T$. Actually,
when $f(t)\ll T_2$, we can expand $k_V(\frac{2T_2+ f(t)}{2})$ into Taylor series and
obtain the leading term as $\mean{J}\approx{k'_V(T_2)}\overline{f^2(t)}/{8\gamma}$, where
$'$ denotes derivative and $\overline{\cdots}$ denotes time average over one period. This
pump effect is more significant if $k'_V(T_2)$ is larger; and is vanishing when
$k'_V(T_2)=0$.

Since the effective force constant is independent of $T$ for harmonic potential, the
phenomena mentioned above cannot be observed using harmonic potential.
On the other hand, is harmonic potential the only potential that has such a property? The
answer is no. An example is the Toda potential $V= k(e^{-ax}+ ax)$ for which
$\meant{V''}= k a^2$.
Interestingly, both the harmonic lattice and Toda lattice are
models displaying ballistic heat transport. These facts imply a clue that
temperature independent $k_V(T)$ is related to ballistic transport.

\begin{figure}[t!]
\centering
 \includegraphics[width=0.35\textwidth]{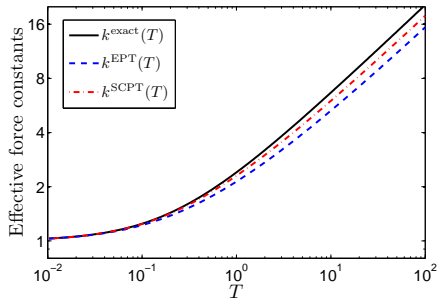}
 \caption{Effective force constant from different theories for the FPU-$\beta$ model with
$k_2=1$ and $k_4=1$.}
\label{fig:comp}
\end{figure}

It is worth mentioning that there are other approximated mean--field theories trying to
convert an anharmonic potential in lattices to effective harmonic potential with
temperature dependent force constant. For example, for the FPU-$\beta$ model
with $V=\frac{k_2}{2}y^2+\frac{k_4}{4} y^4$, the effective phonon theory (EPT)
 estimates it as
$
	k^{\mathrm{EPT}}(T)= k_2+ k_4\fracc{\meant{y^4}}{\meant{y^2}}
$
 \cite{EL.06.Li,PRL.10.Li} while the self-consistent phonon theory (SCPT)
estimates it as
$
k^{\mathrm{SCPT}}(T)= \frac{1}{2}\bkb{k_2+ \sqrt{k_2^2+ 12k_4
T}}
$
\cite{PRE.08.He}.
In Fig.~\ref{fig:comp}, we compare these results with the present exact result for the
two-oscillator model, \ie{}, $k^\mathrm{exact}(T)=k_2+
3k_4  \meant{y^2}$. It is shown that both the EPT and SCPT underestimate the nonlinear
effect,
although at low temperature the deviations are tiny. At the high temperature regime,
 all these theories result in the $\sim c T^{1/2}$ dependence. For $k_2=k_4=1$, the
coefficient $c$ is $6\Gamma(3/4)/\Gamma(1/4)\approx 2.028$ for the exact value. For the
EPT $c=\Gamma(1/4)/2\Gamma(3/4)\approx 1.479$ and for the SCPT $c=\sqrt{3}\approx 1.732$,
which are 27\% and 14\% less than the exact value, respectively.

Upto now we only discussed the first cumulant, \ie{}, the average heat flux.
However, using the standard perturbation theory, all higher order cumulants can be derived.
For example, the second cumulant reads (see for example \cite{NULL.96.Szabo})
\begin{equation}
\label{eq:sec}
	\lim_{t\to\infty}\frac{\mmean{Q^2}}{t}=
	2L^{(2)}_{00}	+2
\sum_{m=1}^{\infty}\frac{
L^{(1)}_{0m}L^{(1)}_{m0}
}{\mu_m-\mu_0},
\end{equation}
where $\mathcal{A}_{mn}\equiv \inner{\ttdvp_m}{\mathcal{A}\ttdp_n}$ for short. Note that
all the terms in the formula are again evaluated at $\lambda=0$.

The excited states are involved in the expression for higher order cumulants.
Therefore, to analytically
obtain the higher order cumulants, we should be able to completely solve all the
eigenstates of the Schr\"odinger equation \eeqref{eq:sdg} with $\lambda=0$, which
describes a single particle with mass $T$ moving in the potential $V_S= V'^2/4T- V''/2$
(see Fig.~\ref{fig:1}(c)).
This is of course an impossible mission for most of the anharmonic potentials. However,
numerically solving it by diagonalizing the operator $L_S$ is straightforward. In this
sense, our method provides a new numerical approach other than molecular dynamics to
investigate the nonlinear heat transport problems.

It can also be noticed that,
even we are not able to obtain the CGF $\tdg=-\tdm_0 t$ as a function of $\lambda$ for
general $V$, we can still verify the GC symmetry
\begin{equation}
	\tdg(\lambda)=\tdg(\Delta \beta-\lambda),
\end{equation}
where $\Delta\beta= 1/T_1-1/T_2$. This result can be seen from Eqs. (\ref{eq:sdg}) and
(\ref{eq:sdg2}). Because $\tdv$ possess this symmetry, all its eigenvalues $\tdm_n$ and
eigenfunctions $\tdps_n$ should possess this symmetry. \mytextcolor{blue}{However,
$\tdp_n$ and $\tdvp_n$
do not follow it because the transformation $e^{\pm V/2\tilde{T}}$ contains $\tilde{T}$
which is not invariant and hence GC symmetry is not satisfied in the transient stage.
It is only valid in the long-time limit when the largest eigenvalue dominates.}

\subsection{Geometric Heat Fluxes\label{sec:geom}}
It has been discovered that in classic heat transport, the geometric curvature might
contribute to heat transport when two or
more system parameters are modulated adiabatically \cite{PRL.12.Ren}.

When two parameters $u_1$ and $u_2$ are
modulated, the average geometric heat flux reads
\begin{equation}
	\mean{J}_{geom}= -\frac{1}{T_p}\iint_{u_1u_2}
\dif{\tilde{\mathcal{F}}_{u_1u_2}} du_1du_2,
\end{equation}
where
$\tilde{\mathcal{F}}_{u_1u_2}$ is the geometric
curvature in the $(u_1, u_2)$ space
\begin{equation}
\label{eq:curve}
	\tilde{\mathcal{F}}_{u_1u_2}=
\inner{\lpt{\tdvp_0}{u_1}}{\lpt{\tdp_0}{u_2}}-\inner{\lpt{\tdvp_0}{u_2}}{\lpt{\tdp_0}{u_1}
}.
\end{equation}
In the expression, $\tdvp_0$ and $\tdp_0$ should satisfy the normalization condition
\eeqref{eq:norm}. However, we are free of using different conventions to normalize these
eigenfunctions.
It can be shown that the curvature shown above will not be affected by how we
normalize the eigenfunctions, which is the analogue to ``gauge invariance'' in quantum
mechanics when dealing with Berry curvature.

Typically, in quantum mechanics, to normalize an unnormalized eigenfunction
$\tdps_n$, we need to choose $\tdps_n\to \inner{\tdps_n}{\tdps_n}^{-1/2}
\tdps_n$. 
For the Fokker-Planck
equation discussed here, we can still use this method and then transform $\tdps_n$ to
$\tdvp_n (\tdp_n)=
e^{\pm V/2\tilde{T}}\tdps_n$. But this will make the later derivation difficult. A more
convenient way in our situation is to choose
\begin{equation}
\ttdp_0(y)=\Pst= \frac{1}{Z_T}e^{-\frac{V(y)}{T}} \mathsp{and} \ttdvp_0(y)=1.
\end{equation}
In other words, $\ttdp_0(y)$ is the Boltzmann distribution and $\ttdvp_0(y)$ is a
constant. We call this the Boltzmann convention.
In this convention, it can be derived that
\begin{equation}
\label{eq:f0}
	\ddd{\tilde{\mathcal{F}}}_{u_1u_2}=\inner{\lpt{\ddd\tdvp_0}{u_1}}{\lpt{ \ttdp_0} {
u_2} }
	-\inner{\lpt{\ddd\tdvp_0}{u_2}}{\lpt{\ttdp_0}{u_1}},
\end{equation}
in which for any quantity $\mathcal{A}$, we use the notation
\begin{equation}
\label{eq:curve0}
	\ddd{\mathcal{\tilde A}}\equiv \dif{\mathcal{\tilde A}}.
\end{equation}

\subsubsection{Modulating \texorpdfstring{$T_1$}{T1} and \texorpdfstring{$T_2$}{T2}}

It has been discovered that when the
two bath temperatures are modulated, there is no geometric contribution to the average
heat flux for the harmonic potential. However, for the anharmonic FPU-$\beta$ model, this
effect has been observed numerically \cite{PRL.12.Ren}.

To speculate the role of anharmonic potential $V$ on geometric
heat flux when $T_1$ and $T_2$ are modulated, we derive the following equation for the
curvature (see Appendix \ref{sec:app})
\begin{equation}
\label{eq:geom}
	\ddd{{\tilde{\mathcal{F}}_{T_1T_2}}}=
	\lptl{{\tilde{\mathcal{F}}_{T_1T_2}}}\Big|_{\lambda=0}
	=\frac{1}{T^2(\gamma_1+\gamma_2)}
\sum_{
m=1} ^{ \infty}
\frac{{V''}_{0m}V_{m0}}{\mu_m-\mu_0},
\end{equation}
where we still use
$\mathcal{A}_{mn}\equiv\inner{\ttdvp_m}{\mathcal{A}\ttdp_n}$ for any $\mathcal{A}$.
However,
since both $V''$ and $V$ are functions of $y$ instead of operators, here we can also
evaluate
them as
$\mathcal{A}_{mn}=\inner{\Psi_m}{\mathcal{A}\Psi_n}=\mathcal{A}_{nm}$.

Now we are able to
distinguish the effect of anharmonicity to the geometric heat flux. For harmonic potential
$V=\frac{1}{2}ky^2$, ${V''}_{0m}= k\inner{\Psi_0}{\Psi_m}$, which is obviously zero
for $m\ne 0$. Therefore, at any point $(T_1, T_2)$ in the parameter space,
$\dif{\tilde{\mathcal{F}}_{T_1T_2}}=0$ and finally $\mean{J}_{geom}=0$. For anharmonic
potentials, this term would generally be nonzero, which results in a non-vanishing
curvature and geometric heat flux.

Similar to \eeqref{eq:sec}, it can be seen that only when we can obtain all the excited
states, are we able to get the final result of \eeqref{eq:geom}.
However, for weak nonlinear cases, \eeqref{eq:sec} and
(\ref{eq:geom}) can be used to obtain the approximated values. Consider the
case
$V(y)=\frac{1}{2}k_2 y^2+ \alpha U(y)$, with a small $\alpha$. By regarding $\alpha
U(y)$ as the perturbation, we can evaluate \eeqref{eq:geom} to the first order of
$\alpha$ as
\begin{equation}
\label{eq:pert}
	\ddd{{\tilde{\mathcal{F}}_{T_1T_2}}}\!\approx\! \alpha\frac{}{}
\frac{\inner{\ttdps_0^{(0)}}{U''\ttdps_2^{(0)}}
\inner{\ttdps_2^{(0)}}{\frac{1}{2}k y^2
\ttdps_0^{(0)}}}{\mu_2^{(0)}(\gamma_1+\gamma_2)T^2},
\end{equation}
where $(0)$ in the superscript indicates the unperturbed correspondences, which are for
the harmonic potential. For the FPU-$\beta$ model with $U(y)=k_4 y^4/4$, \eeqref{eq:pert}
becomes $\frac{3\gamma k_4}{2(\gamma_1+\gamma_2)k_2^2}\alpha$, which is independent of the
temperature. Therefore, the heat transfered in one period should be proportional to the
area enclosed by the driving path.

\subsubsection{Modulating \texorpdfstring{$\gamma_1$}{gamma1} and
\texorpdfstring{$\gamma_2$}{gamma2}}
If we fix $T_1=T_2=T_0$ and module $u_{1,2}= \gamma_{1,2}$,
then we cannot obtain a geometric heat flux, for any $V(y)$. This is because in
this case $T=T_0$, which is independent of $\gamma_1$ and $\gamma_2$. As a consequence,
\begin{equation}
	\ttdp_0= \frac{e^{-V(y)/T}}{\int e^{-V(y)/T} dy}
\end{equation}
is also independent of $\gamma_1$ and $\gamma_2$. Inserting this result into
\eeqref{eq:f0}
immediately shows $\ddd{\tilde{\mathcal{F}}}_{\gamma_1\gamma_2}=0$ and therefore
$\mean{J}_{geom}=0$.

The same result has also been found in
\cite{PRL.10.Ren} for quantum heat transport and in \cite{PRL.12.Ren} for classic case
with
harmonic potential.

\subsection{An Example} Next, we discuss an anharmonic model that is analytically
solvable, that is Model I with a V-shaped potential $V(y)= k|y|$.
Correspondingly,
${V}_S=k^2/4T -k\delta(y)$, which is an
infinitely deep well potential plus a constant shift.
This shift will only lift the eigenvalues and will not affect the eigenstates. Therefore,
the full spectrum of the Schr\"odinger equation with $\lambda=0$, as well
as all the cumulants, should be solvable.

Actually, for this potential, we can directly solve
\eeqref{eq:sdg}.
It is easy to show that only the ground state is bounded and the
corresponding eigenvalue is
\begin{equation}
	\tdm_0
	= -\frac{k^2T_1T_2}{T(\gamma_1+\gamma_2)}
\bks{\lambda- \left(\frac{1}{T_1}-\frac{1}{T_2}\right)}\lambda,
\end{equation}
which is a quadratic form. Therefore, in the long-time limit, $Q$ will be a Gaussian
random variable, which only has the lowest two  cumulants:
$	\lim_{t\to\infty} \frac{\mmean{Q}}{t}=
\frac{k^2}{(\gamma_1+\gamma_2)T}(T_1-T_2) =\mean{J}$ and
$	\lim_{t\to\infty} \frac{\mmean{Q^2}}{t}= \frac{2k^2 T_1 T_2}{ T(\gamma_1
+\gamma_2)}.
$

To obtain the geometric heat flux, the ground state eigenfunctions are needed, which read
\begin{align}
	\tdvp_0&= e^{-k|y|
\frac{\lambda T_1/\gamma_1}{T_1/\gamma_1+T_2/\gamma_2}
}
;\\ \tdp_0&= \frac{k}{2T} e^{-k|y|
\frac{1/\gamma_1+ 1/\gamma_2-\lambda T_1/\gamma_1}{T_1/\gamma_1+T_2/\gamma_2}
}.
\end{align}
It is then easy to obtain $
	\tilde{\mathcal{F}}_{T_1T_2}= -\frac{\gamma}{\gamma_1+\gamma_2}\frac{\lambda}{T}.
$
Interestingly, this curvature is proportional to $\lambda$. Therefore, the first
cumulant
is nonzero and all higher order cumulants are zero.


\section{\label{sec:md2}Model II} The idea presented above is not restricted to the
overdamped two-oscillator model, but generally applicable to all heat transport problems.
If the steady state can be solved, then at least the average heat flux can be obtained.
This is the situation also for Model II, a bounded single-oscillator model with inertia,
following
the
Langevin dynamics:
\begin{align}
	\dot{y}&=u; \\
	m \dot{u}&=- V'(y)-\gamma_1 u-\gamma_2 u+ \xi_1+ \xi_2.
\end{align}
Meanwhile, the heat $Q$ satisfies
\begin{equation}
	\dot{Q}= (-\gamma_1 \dot{y}+ \xi_1)\dot{y}
\end{equation}
and their joint probability $\tilde P(y,u)$ satisfies the Fokker-Planck equation
\begin{equation}
\begin{split}
	&\lpt{}{t}\tilde P(y,u)=\tdl \tilde P(y,u)\\
	&= (L^{(0)}+ \lambda L^{(1)}+ \lambda^2 L^{(2)})\tilde
P(y,u)\\
	&=
	\left[
-\lpt{}{y}u+\lpt{}{u}\bkb{\frac{\gamma}{m}u+\frac{V'(y)}{m}}
+\lppt{2}{}{u}\frac{\gamma T}{m^2}\right. \\
&\left.-\lambda \bkb{\gamma_1 u^2-\frac{\gamma_1 T_1}{m}+ 2\lpt{}{u}\frac{\gamma_1 T_1}{m}
u}
+\lambda^2 \gamma_1 T_1 u^2\right] \tilde P(y,u),
\end{split}
\label{eq:md2}
\end{equation}
where $T=(\gamma_1 T_1 + \gamma_2 T_2 )/(\gamma_1 +
\gamma_2)$ and
$\gamma=\gamma_1+\gamma_2$.

\newcommand{\hf}{\frac{1}{2}}
For $\lambda=0$, \eeqref{eq:md2} is a single particle Kramers equation. It has a unique
steady state (ground state), which is just the Boltzmann distribution
\cite{NULL.98.Risken}
\begin{equation}
\label{eq:boltz2}
P_{T}^{\mathrm st}(y,u)=\frac{1}{Z} e^{-H/T} \quad \mathrm{with}\quad H=\hf mu^2+ V(y),
\end{equation}
and $Z$ being the partition function.

Therefore, we can directly use our formalism to obtain the dynamic heat flux as
\begin{equation}
  \bar J= \iint L^{(1)} P_T^{\mathrm st} dydu= \frac{\gamma_1 \gamma_2}{m
(\gamma_1+\gamma_2)} (T_1-T_2),
\end{equation}
which is independent of $V(y)$ because $L^{(1)}$ is independent of $y$.
The immediate question that can be
raised is whether
the higher cumulants are also independent of $V(y)$.

In fact, for this model, the ground state of \eeqref{eq:md2}, {\it i.e.},
\begin{equation}
  \tdl\tdp_0(y,u)\!=\!(L^{(0)}+ \lambda L^{(1)}+ \lambda^2 L^{(2)})\tdp_0(y,u)\!=\!
-\tdm_0
\tdp_0(y,u),
\end{equation}
as well as the left eigen problem
\begin{equation}
 \tdl^\dag \tdvp_0(y,u)= -\tdm_0 \tdvp_0(y,u)
\end{equation}
 can be solved using trial solutions
\newcommand{\ttt}{\tilde{\mathcal{T}}}
\begin{equation}
\label{eq:lr}
 {\tdvp}_0 (y,u) \propto e^{-H/\ttt_L}\mathsp{and} {\tdp}_0 (y,u) \propto e^{-H/\ttt_R},
\end{equation}
in which $\ttt_{L,R}$ is independent of $y$ and $u$.

For the right eigenfunction and eigenvalue, we get
\begin{equation}
  \tdl \tdp_0 = [g(\ttt_R)+ h(\ttt_R) u^2] \tdp_0,
\end{equation}
where
\begin{align}
	g(\ttt_R)&= \frac{\gamma (\ttt_R-T)-\gamma_1 \lambda  T_1\ttt_R}{m \ttt_R}; \\
	h(\ttt_R)&= \frac{\gamma_1 \lambda(\lambda T_1 -1)\ttt_R^2+(2
\gamma_1\lambda T_1 - \gamma)\ttt_R+\gamma T}{\ttt_R^2}.
\end{align}
To make it an eigenvalue problem, we demand $h(\ttt_R)=0$. Solving it for
$\ttt_R$ gives
\begin{equation}
	\ttt_R\!=\! \frac{\gamma\!-\!2\gamma_1 \lambda T_1\! \pm\!
	\sqrt{
	(\gamma-2\gamma_1\lambda T_1)^2\!-\! 4\gamma\gamma_1 \lambda T(\lambda T_1-1)
	}
	}{2\gamma_1 \lambda(\lambda  T_1\!-\!
1)}.
\end{equation}
There are two possible solutions. However, we know that the ground state should converge
to the Boltzmann distribution \eeqref{eq:boltz2}. Therefore, it is required that
\begin{equation}
	\lim_{\lambda\to 0} \ttt_R= T=\frac{\gamma_1 T_1+\gamma_2 T_2}{\gamma_1+\gamma_2}.
\end{equation}
So we need choose the solution with minus ``-'' sign.
Consequently, we obtain the corresponding eigenvalue
\begin{equation}
\begin{split}
	\tdm_0&=-g(\ttt_R)  \\
	&=\!\frac{\gamma_1\!+\!\gamma_2}{2m}\!\bks{
\sqrt{1\!+\!\frac{4\gamma_1\gamma_2}{(\gamma_1\!+\!\gamma_2)^2}T_1T_2\lambda
\left(\frac{1} { T_1 }\!-\!\frac{1}{T_2}\!-\!\lambda\right)
}\!-\!1
},
\end{split}
\end{equation}
which is indeed independent of $V$. Therefore, all the higher order cumulants are also
independent of $V$ at long time. Previously, this formula has been obtained for
the harmonic case \cite{JSM.11.Kundu,JSM.11.Fogedby} and
numerically tested for the FPU-$\beta$ potential \cite{JSM.11.Fogedby}.

Similarly, the left ground state eigenfunction can be obtained, which is in the form
of \eeqref{eq:lr}
with
\begin{equation}
\ttt_L\!=\! \frac{2\gamma_1 \lambda T_1\!-\!\gamma\! -\!
	\sqrt{
	(\gamma-2\gamma_1\lambda T_1)^2\!-\! 4\gamma\gamma_1 \lambda T(\lambda T_1-1)
	}
	}{2\gamma_1 \lambda(\lambda  T_1\!-\!
1)},
\end{equation}
since the left eigenfunction for $\lambda=0$ should be 1 ($\ttt_L\to \infty$).

With the ground state eigenvalue and eigenfunctions known, we can easily calculate other
quantities as well, such as the geometric curvature, which is left to the readers, if
interested.

\section{\label{sec:summary}Summary}
Using the Fokker-Planck equation for the full counting statistics for heat, we have
developed
a formalism to obtain the dynamic and geometric heat current for arbitrary anharmonic potential. The importance of
anharmonicity in manipulating heat is investigated and it is found that the temperature dependent force constant plays
a key role. The higher cumulants can also be obtained if the eigenspectrum of the original
FP operator $L^{(0)}$ is known. As an example, we choose
$V$ shaped potential and  obtained all the cumulants explicitly. We also extend this method by including the inertia term for a bounded
single particle and found that the cumulants are independent of the nonlinear onsite
potential. For arbitrary nonlinear potential, the Gallavotti-Cohen symmetry is verified
for both models.

In common sense, the nonlinear transport problems are generally thought to be not exactly
solvable. We hope the existance of exact solutions, presented here, for nonlinear
transport problems will motivate further investigations on solving more general anharmonic
systems, such as heat transport in nonlinear chain models which is one of the ultimate
goals in the community of heat transport.

\acknowledgements {This work is supported in part by the MOE T2 Grant R-144-000-300-112
and an NUS Grant
R-144-000-285-646.}

\appendix

\section{Derivation of \eeqref{eq:geom}}
\label{sec:app}
\newcommand{\mA}{\mathcal{A}}
\newcommand{\vvv}[2]{\big(#1\big)_{T_{#2}}}
\newcommand{\vvt}[3]{\vvv{\lpt{#1}{T_{#2}}}{#3}}
\newcommand{\df}[2]{\frac{d #1}{d #2}}
\newcommand{\vvtt}[3]{\df{#1}{T}\vvv{\lpt{T}{T_{#2}}}{#3}}
\newcommand{\ttdm}{{\mu}}
In this appendix, we derive \eeqref{eq:geom} from \eeqref{eq:f0} with
$u_{1,2}=T_{1,2}$.

We first note that $\ddd\tdvp_0$ can be obtained using first order perturbation
theory, which reads
\begin{equation}
\label{eq:tdvp}
 \ddd\tdvp_0= \sum_{m=1}^{\infty}
\frac{(\ttdvp_0,{L^{(1)}} \ttdp_m)}{\ttdm_m-\ttdm_0}
\ttdvp_m + a_{00} \ttdvp_0,
\end{equation}
where $a_{00}$ depends on the normalization convention used.
However, from \eeqref{eq:curve}, it can be easily
shown that the projection of $\ddd\tdvp_0(\ddd\tdp_0)$ on $\ttdvp_0(\ttdp_0)$ will not
contribute to
$\ddd{\mathcal{F}}_{u_1u_2}=\partial_\lambda\mathcal{F}_{u_1u_2}|_{\lambda=0} $ for any
$u_{1,2}$. Then we can just omit this term and regard $a_{00}=0$.

Then it can be noticed that all the eigenfunctions $\ttdvp_m$, $\ttdp_m$, as well as the
eigenvalues
$\mu_m$, are
explicit functions of $T$. Their $T_{1,2}$-dependence comes from
$T(T_1,T_2)= (\gamma_2 T_1+ \gamma_1 T_2)/(\gamma_1+\gamma_2)$. Therefore, for $\mA=
\ttdvp_m, \ttdp_m$ or
$\mu_m$,
\begin{equation}
	\vvv{\lpt{\mA}{T_1}}{2}= \vvv{\lpt{T}{T_1}}{2}\df{\mA}{T}=
\frac{\gamma_2}{\gamma_1+\gamma_2}
	\df{\mA}{T}\mathsp{and} 1\longleftrightarrow 2,
\end{equation}
where $\vvv{\cdot}{2}$ means partial derivative while keeping $T_2$ invariant. Meanwhile,
$L^{(1)}$ is an explicit function of $T_1$ only,
\begin{equation}
	L_A \equiv \vvt{L^{(1)}}{1}{2}=
\frac{V''}{\gamma_1}-\frac{2}{\gamma_1}\lpt{}{y}V'\mathsp{and}
	\vvt{L^{(1)}}{2}{1}= 0.
\end{equation}
Therefore, we can regard $\ddd\tdvp_0$ as an explicit function of $T$ and $T_1$. From the
chain rule we get
\begin{equation}
	\vvt{\ddd\tdvp_0}{1}{2}= \vvt{\ddd\tdvp_0}{1}{}+ \vvt{\ddd\tdvp_0}{}{1}\vvt{T}{1}{2},
\end{equation}
and
\begin{equation}
	\vvt{\ddd\tdvp_0}{2}{1}= \vvt{\ddd\tdvp_0}{}{1}\vvt{T}{2}{1}.
\end{equation}

Therefore, in the Boltzmann convention,
\begin{equation}
\label{eq:f01}
	\begin{split}
		\ddd{\tilde{\mathcal{F}}}_{T_1T_2}=& \inner{\vvt{\ddd\tdvp_0}{1}{2}}{\vvt{
\ttdp_0} {2}{1} }
	-\inner{\vvt{\ddd\tdvp_0}{2}{1}}{\vvt{\ttdp_0}{1}{2}} \\
		=&  \inner{\vvt{\ddd\tdvp_0}{1}{}+ \vvt{\ddd\tdvp_0}{}{1}\vvt{T}{1}{2}
		}{\vvtt{ \ttdp_0} {2}{1}  }\\
		&-\inner{ \vvt{\ddd\tdvp_0}{}{1}\vvt{T}{2}{1}}{\vvtt{\ttdp_0}{1}{2}} \\
		=& \inner{\vvt{\ddd\tdvp_0}{1}{}}{\vvtt{ \ttdp_0} {2}{1}}.
	\end{split}
\end{equation}
Noticing that in
\eeqref{eq:tdvp}, only $L^{(1)}$ explicitly contains $T_1$, we get (omit the term
$a_{00}\varphi_0$)
\begin{equation}
\label{eq:f1}
	\vvt{\ddd\tdvp_0}{1}{}\!=\! \sum_{m=1}^{\infty}
\frac{(\ttdvp_0,\vvt{L^{(1)}}{1}{} \ttdp_m)}{\ttdm_m-\ttdm_0}
\ttdvp_m
\!=\! \sum_{m=1}^{\infty}
\frac{\inner{\ttdvp_0}{L_A\ttdp_m}}{\ttdm_m-\ttdm_0}
\ttdvp_m.
\end{equation}
In the Boltzmann convention, $\ttdvp_0=1$, so that
\begin{equation}
\label{eq:f2}
\begin{split}
	\inner{\ttdvp_0}{L_A\ttdp_m}&= \int L_A\ttdp_m(y)dy \\
	&= \int \frac{V''}{\gamma_1} \ttdp_m(y)dy
	= \frac{\inner{\ttdvp_0}{V''
\ttdp_m}}{\gamma_1}
\end{split}
\end{equation}
due to the natural boundary conditions.

Finally, inserting Eqs.~(\ref{eq:f1}) and (\ref{eq:f2}) into \eeqref{eq:f01}, and noticing
	\begin{equation}
		\df{\ttdp_0}{T}=\df{}{T}\frac{e^{- V/T}}{Z_T}
= \frac{V- \meant{V}}{T^2} \ttdp_0,
	\end{equation}
we obtain
\begin{equation}
\begin{split}
	\ddd{\tilde{\mathcal{F}}}_{T_1T_2}&= \sum_{m=1}^{\infty}
\frac{\inner{\ttdvp_0}{\frac{V''}{\gamma_1}\ttdp_m}}{\ttdm_m-\ttdm_0}
\inner{\ttdvp_m}{\frac{V- \meant{V}}{T^2} \ttdp_0}\frac{\gamma_1}{\gamma_1+\gamma_2} \\
&=\frac{1}{T^2(\gamma_1+\gamma_2)}\sum_{m=1}^{\infty}
\frac{
\inner{\ttdvp_0}{V''\ttdp_m}\inner{\ttdvp_m}{V\ttdp_0}}
{\ttdm_m-\ttdm_0},
\end{split}
\end{equation}
where $\inner{\ttdvp_m}{\meant{V}\ttdp_0}=0$ for $m\ne 1$ is used.


\bibliography{a}
\end{document}